\documentclass[twocolumn]{article}\usepackage[]{graphicx}\usepackage[]{xcolor}
\makeatletter
\def\maxwidth{ %
  \ifdim\Gin@nat@width>\linewidth
    \linewidth
  \else
    \Gin@nat@width
  \fi
}
\makeatother

\definecolor{fgcolor}{rgb}{0.345, 0.345, 0.345}
\newcommand{\hlkwb}[1]{\textcolor[rgb]{0.69,0.353,0.396}{#1}}%

\usepackage{framed}
\makeatletter
 {\par\unskip\endMakeFramed%
 \at@end@of@kframe}
\makeatother

\definecolor{shadecolor}{rgb}{.97, .97, .97}
\definecolor{messagecolor}{rgb}{0, 0, 0}
\definecolor{warningcolor}{rgb}{1, 0, 1}
\definecolor{errorcolor}{rgb}{1, 0, 0}
\usepackage{alltt}
\usepackage{booktabs}
\usepackage{array}
\usepackage[margin=1in]{geometry}
\usepackage[english]{babel}
\usepackage{amsmath}
\usepackage{todonotes}

\usepackage{hyperref}
\usepackage[numbers]{natbib}
\newcommand{\abs}[1]{\left\lvert#1\right\rvert} % Absolutbetrag

\date{}
\IfFileExists{upquote.sty}{\usepackage{upquote}}{}
\begin{document}

\title{ \huge{\textbf{Addressing Outcome Reporting Bias in Meta-analysis: A Selection Model Perspective}}}
%\author{\textbf{Alessandra Gaia Saracini and Leonhard Held}}
\author{\textbf{Alessandra Gaia Saracini\textsuperscript{1} and Leonhard Held\textsuperscript{2}}}

\twocolumn[
\begin{@twocolumnfalse}
  \maketitle
  \begin{abstract}
  Outcome Reporting Bias (ORB) poses significant threats to the validity of meta-analytic findings. It occurs when researchers selectively report outcomes based on the significance or direction of results, potentially leading to distorted treatment effect estimates. Despite its critical implications, ORB remains an under-recognized issue, with few comprehensive adjustment methods available. The goal of this research is to investigate ORB-adjustment techniques through a selection model lens, thereby extending some of the existing methodological approaches available in the literature. To gain a better insight into the effects of ORB in meta-analysis of clinical trials, specifically in the presence of heterogeneity, and to assess the effectiveness of ORB-adjustment techniques, we apply the methodology to real clinical data affected by ORB and conduct a simulation study focusing on treatment effect estimation with a secondary interest in heterogeneity quantification.\\
  \end{abstract}
\end{@twocolumnfalse}
]

\footnotetext[1]{Corresponding author: Alessandra Gaia Saracini\\
University of Zurich\\
Epidemiology, Biostatistics and Prevention Institute\\
ETH Zurich Department of Mathematics\\
alessandragaia.saracini@gmail.com}
\footnotetext[2]{Leonhard Held, Professor\\
University of Zurich\\
Epidemiology, Biostatistics and Prevention Institute\\
leonhard.held@uzh.ch}

\section{Introduction}
Meta-analysis is a powerful statistical tool used to combine evidence from multiple studies investigating the same research question \citep{DerSimonian, handbook}. It plays a crucial role in clinical research by providing a more comprehensive and robust analysis of treatment effects, especially when individual studies have limited statistical power. However, like any statistical method, meta-analysis is prone to biases that can affect its validity and reliability \citep{handbook, Schmid2022, Egger2022}. While publication bias (PB) is a well-known issue, with various statistical methods developed to address it, outcome reporting bias (ORB) is less explored but equally problematic \citep{handbook, Schmid2022, Egger2022, protocolORB}. PB occurs when entire studies are not present in the literature due to the lack of significance or direction of results. On the other hand, ORB occurs when reporting decisions within published studies are influenced by results' significance or direction, leading to selective reporting of outcomes \citep{ORBimpact, Copas2014, Copas2019, Schmid2022, Kirkham2012, Egger2022, dutch, ORBIT_paper, Thomas2022}. Therefore, unlike PB, studies affected by ORB may still be published, but certain outcomes, especially those with weaker results, may be omitted or reporting may be impartial, leading to inability to include the study outcome in a meta-analysis. 

Studies have shown that ORB is prevalent in the meta-analysis literature, affecting reviews where both primary and secondary outcomes are often inadequately reported \citep{moreORBevidence, ORBimpact, ORBIT_paper, silva_many_2024, Thomas2022}. An investigation on a cohort of Cochrane systematic reviews by \citet{ORBimpact} found that more than half of the reviews did not include full data for the primary outcome of interest from eligible trials, and over a third contained at least one trial with high suspicion of ORB \citep{ORBimpact}. An investigation by \citet{ORBIT_paper}, with a focus on meta-analyses where the primary outcome was a harmful one, found that $86\%$ of Cochrane cohort reviews did not include full outcome data for the main adverse event of the trial, and ORB was suspected in nearly two thirds of the reviews \citep{ORBIT_paper}. Even prospective trial registration, intended to discourage selective reporting and enhance transparency, does not always succeed in preventing ORB. Trial registries often lack statistical analysis plans and are not externally audited, allowing researchers to adapt analyses or selectively report outcomes based on significance \citep{Thomas2022, silva_many_2024}. This limitation is reflected in the findings of \citet{silva_many_2024}, who examined 84 prospectively registered trials and found that nearly half showed evidence of selective outcome reporting, with over $20\%$ omitting a registered primary outcome entirely. A study by \citet{moreORBevidence}, inspecting 1402 outcomes from 48 trials with 68 publications, quantified the association between inadequate reporting of outcomes and statistical significance. They concluded that statistically significant beneficial outcomes have odds of being fully reported which are 2.7 times that of non-significant ones, with a $95 \%$ CI from 1.5 to 5.0 \citep{moreORBevidence}. Similar investigations on RTCs estimated that statistically significant outcomes were 2.2–4.7 times more likely to be fully reported compared with outcomes that were not statistically significant \citep{Thomas2022, Dwan2008}. ORB poses a substantial threat to the integrity of meta-analyses, emphasizing the need for increased awareness and methods to mitigate its impact.

Omitting study outcomes based on their effect size or significance, leading to ORB, can be equivalently understood from a missing data perspective. Through this lens, ORB reflects a missing not at random (MNAR) scenario, meaning that the probability of an outcome being (un)reported depends on its value, or its statistical significance. This contrasts with the assumption of missing completely at random (MCAR), where missingness is unrelated to the data. Standard meta-analytic methods, which only include reported outcomes in an analysis, implicitly assume a MCAR setting, leading to biased estimates when the true mechanism is MNAR \citep{Hwang2018, Liu2017, dutch}.

A related but distinct setting is missing at random (MAR), where the reporting of one outcome depends on other observed outcomes within the same investigation, which can also lead to ORB \citep{dutch, Hwang2018, Liu2017}. This is particularly relevant in ORB-adjustments in multivariate meta-analysis, where methods have been developed to leverage correlations between outcomes and borrow strength across them in case of missing information \citep{Kirkham2012, Bay, dutch, Hwang2018, Liu2017}. Key examples include the bivariate meta-analysis of \citet{Kirkham2012} and its Bayesian extension by \citet{Bay}, as well as expansions to indirect treatment comparisons and network meta-analysis by \citet{Hwang2018} and \citet{Liu2017}. These approaches have shown to reduce bias under MAR, especially when outcomes are strongly correlated. However, estimating the between and within study correlation structures is often difficult. Some methods address this challenge by assuming a global correlation coefficient \citep{Kirkham2012, Bay, dutch, Hwang2018, RileyPG, Riley}, estimated either via Pearson’s correlation \citep{Kirkham2012, dutch}, or, in the case of \citet{Liu2017} using copula models, e.g., the Clayton copula, or via non-informative priors in a Bayesian setting \citep{Bay, Liu2017, Hwang2018}. Although these strategies offer partial improvements under MNAR, they generally do not model the MNAR mechanism directly and may inadvertently introduce bias into the estimated effects of outcomes which are fully observed due to the joint modeling assumptions.

In contrast, the most established MNAR-based ORB-adjustment is the method proposed by \citet{Copas2019, Copas2014}, which adjusts each outcome separately and explicitly models the missing data mechanism. The method relies on first categorizing unreported outcomes into risk categories - no risk (NR), low risk (LR), and high risk (HR) - based on the Outcome Reporting Bias in Trials (ORBIT) classification system. Assuming these categories are correctly assigned, the likelihood-based model includes contributions from unreported HR outcomes, under the assumption that these would have been non-significant. Of note, the ORBIT classification is specific to ORB as it evaluates the risk of bias at the level of each unreported outcome within a study, rather than at the study level. In contrast, study level risk of bias assessments, typical of PB \citep{Lunny2024,RiskBias, Begg, Salanti}, are less suitable for ORB, since published studies may contain multiple unreported outcomes, each with potentially different levels of risks of ORB \citep{Copas2019, Copas2014, Kirkham2012}.

Our work builds on this foundation by reinterpreting ORB-adjustment through the lens of selection models, a framework commonly used for PB adjustment \citep{selection2, selection1, reviewselection}. Our proposed method removes the need for ORBIT classification, allows flexible modeling of the missing data mechanism, and incorporates all unreported outcomes in the analysis, not just the ones classified as HR of bias by the ORBIT methodology. We also investigate the influence of between-study heterogeneity - a novel aspect in the ORB literature - and assess the effectiveness of our method through simulations. The focus of the simulations is on treatment effect estimation, with a secondary emphasis on how heterogeneity interacts with ORB and its correction. 

%Our work can be seen as an extension of the \citet{Copas2019} method by presenting ORB-adjustment through a selection model perspective, a framework typically used for PB adjustment \citep{selection2, selection1, reviewselection}. The proposed approach for ORB adjustment offers a more flexible framework that does not require the ORBIT classification system, includes contributions from all unreported study outcomes, and allows for different assumptions on the missing data mechanism. We further consider the impact of between-study heterogeneity on ORB and ORB-adjustment, a novel aspect in the context of ORB, and conduct a simulation study investigating the impact of ORB and the effectiveness of ORB-adjustment, focusing on treatment effect estimation, with a secondary focus on heterogeneity, under different meta-analytic settings. 

Throughout this work, we consider a random effects meta-analysis setting on a single beneficial outcome, i.e., an outcome for which a positive value indicates a beneficial direction of treatment. We assume normality and hence the following model:

\bigskip

\begin{equation}\label{eq:random.eff11}
y_i \sim \mathcal{N}(\theta_i, \sigma_i^2) \; \; \; \; \; \; \theta_i \sim \mathcal{N}(\mu, \tau^2),
\end{equation}

\bigskip

where $y_i$ and $\sigma_i^2$ are the observed treatment effect and squared standard error, respectively, for each study $i$, and the parameters of interest are the treatment effect $\mu$ and the heterogeneity variance $\tau^2$.

As a motivating example of ORB in meta-analysis, we consider the data used by \citet{Copas2019}, wherein a meta-analysis of 12 studies was conducted separately for 14 different outcomes, 2 considered beneficial and 12 harmful. The meta-analysis, originally by \citet{topiramate}, includes studies investigating the effect of Topiramate, an antiepileptic drug first marketed in 1996, when used as an add-on treatment for drug-resistant focal epilepsy. Given that in our research we focus on ORB and ORB correction for beneficial outcomes, we consider the 2 outcomes of the data assumed to have a positive effect, i.e., $50 \%$ seizure frequency reduction, and seizure freedom, illustrated in Table \ref{tab}. We observe that all of the 12 studies in the meta-analysis report the treatment and control arm sample sizes; however, some studies do not report the event frequencies, from which the log risk ratio (RR) is computed and used as the normally distributed treatment effect in \eqref{eq:random.eff11}, using a continuity correction in case of empty cell counts \citep{Copas2019}.

\begin{table}[!h]
\centering
\caption{Results for the outcomes "$50\%$ Seizure Reduction" and "Seizure Freedom" from 12 studies investigating the effect of Topiramate as an add-on treatment (T) vs. control (C) for drug-resistant focal epilepsy \citep{Copas2019, topiramate}.\\[0.5em]}
\resizebox{\linewidth}{!}{
\begin{tabular}{lcccccc}
\toprule
\multicolumn{1}{c}{Study ID} & \multicolumn{2}{c}{Sample Size} & \multicolumn{2}{c}{50\% Seizure Reduction} & \multicolumn{2}{c}{Seizure Freedom} \\
\multicolumn{1}{c}{(Author, date of publication)} & \multicolumn{1}{c}{T} & \multicolumn{1}{c}{C} & \multicolumn{1}{c}{T} & \multicolumn{1}{c}{C} & \multicolumn{1}{c}{T} & \multicolumn{1}{c}{C} \\
%  & 1 & 2 & 3 & 4 & 5 & 6 \\
\midrule
Ben-Menachem 1996 & 28 & 28 & 12 & 0 & Unrep & Unrep \\
Elterman 1999 & 41 & 45 & 16 & 9 & 4 & 2 \\
Faught 1996 & 136 & 45 & 54 & 8 & Unrep & Unrep \\
Guberman 2002 & 171 & 92 & 77 & 22 & 10 & 2 \\
Korean 1999 & 91 & 86 & 45 & 11 & 7 & 1 \\
%\addlinespace
Privitera 1996 & 143 & 47 & 58 & 4 & Unrep & Unrep \\
Rosenfeld 1996 & 167 & 42 & 86 & 8 & Unrep & Unrep \\
Sharief 1996 & 23 & 24 & 8 & 2 & 2 & 0 \\
Tassinari 1996 & 30 & 30 & 14 & 3 & 0 & 0 \\
Yen 2000 & 23 & 23 & 11 & 3 & Unrep & Unrep \\
%\addlinespace
Zhang 2011 & 46 & 40 & 22 & 3 & 0 & 0 \\
Coles 1999 & 52 & 51 & Unrep & Unrep & Unrep & Unrep \\
\bottomrule
\end{tabular}}
\label{tab}
\end{table}

\bigskip

This manuscript is organized as follows: Section \ref{selmodORB} introduces the selection model framework typically used for PB and illustrates the derivations done to adapt this framework to address ORB. In subsection \ref{ourSELs} we elaborate on the various missing data mechanisms considered in our investigations and used in our simulation study, inspired by the PB literature and adapted to ORB. Section \ref{simstudy} presents a simulation study investigating the impact of ORB and the effectiveness of the proposed ORB-adjustment method, with a focus on its application within a random effects meta-analysis model. Subsection \ref{3.1}, \ref{3.2}, and \ref{3.3} describe the simulation setting, key results and the application to the epilepsy example from Table \ref{tab}, respectively. Finally, Section \ref{disc} summarizes the proposed methodology and simulation study findings in a discussion, including limitations and conclusions.

\section{Selection Models for ORB} \label{selmodORB}
Selection models have gained popularity in the publication bias (PB) adjustment literature \citep{DerSimonian, selection1, selection2, reviewselection, handbook}, as they aim at correcting for the bias in treatment effect estimation by directly modelling the assumed missing data mechanism. Let $y_i$ be the observed treatment effect estimate for study $i$ in the meta-analysis, with distribution $f(y_i \text{; } \theta)$, assumed to be normal, where we denote $\theta$ as the unknown parameter of interest - in the context of the random effects meta-analysis of \eqref{eq:random.eff11}, $\theta$ is $\mu$ and $\tau^2$.

The general form of a selection model in the PB literature involves the use of a weighted likelihood function which takes into account the observations $y_i$ from published studies $i \in \{\text{Pub} \}$ by weighting them with a selection function $w(y_i)$ which describes the probability that study $i$ is published/selected based on its significance \citep{selection1, selection2, reviewselection}. By using the following relation:

%\bigskip

\begin{equation}
\label{PB.selection11}
\begin{aligned}
f \left( y_i \text{; } \theta \mid i \in \, \{\text{Pub}\} \right) %\\[0.5em] 
= \frac{f(y_i \text{; } \theta) \cdot w(y_i)}{\int^{+\infty}_{-\infty} f(y \text{; } \theta) \cdot w(y) dy,}
\end{aligned}
\end{equation}

%\bigskip

the PB-adjusted log-likelihood $\ell_{\text{Adj}}^{\text{PB}} \left( \theta  \right)$ is derived \citep{selection0, HedgesVev, selection1, selection2} as

%. The log-likelihood to be maximized for the parameter $\theta$ in turn has an additional contribution for the published studies, $ \{i \in \text{Published} \}$, wherein $f_i(y)$ is weighted by $w_i(y)$, the selection function representing the probability of publishing.

%\bigskip

\begin{equation*}
\begin{aligned}
\label{lik.PB}
\ell_{\text{Adj}}^{\text{PB}}\left(\theta \right) &= \sum_{i} \log f \left(y_i \text{; } \theta \mid i \in \, \{\text{Pub}\} \right)\\
& = \sum_{i \in \{\operatorname{Pub}\}} \log f(y_i \text{; } \theta) \\
& - \sum_{i \in \{\operatorname{Pub}\}} \log \left[ \int_{-\infty}^{\infty} f(y \text{; } \theta) \cdot w(y)  d y \right] \text{.}
\end{aligned}
\end{equation*}

%\bigskip
The selection function can take various forms in the context of PB, generally guided by the intuition that in a meta-analysis of a beneficial outcome, for larger $p$-values, the probability of publication/selection decreases \citep{selection0, HedgesVev, selection1, selection2, selectionWeird, selectionCont, reviewselection}. Of note, in the case of a meta-analysis of a harmful outcome, we expect the opposite: small, significant $p$-values are less likely to be reported, as they would indicate harm \citep{ORBIT_paper, Copas2019}. In the following sections, we define the selection functions assuming beneficial outcomes and thus a positive treatment direction.

The selection function $w(y_i)$ in \eqref{PB.selection11} is therefore often defined as a function of the $p$-value $p_i$, providing an intuitive way of understanding the relationship between significance and the probability of selection \citep{selection1, selection2, reviewselection}. Given that the $p$-value is simply a transformation of the observed treatment effect $y_i$ and the standard error $\sigma_i$, we will use the notation $w(p_i)$ when providing definitions of the selection function in terms of the $p$-value.

%\subsection{Selection Models for ORB} \label{selmodORB}

In the PB selection model setting one takes into account only the non-missing studies $i \in \{\text{Pub} \}$ by defining the conditional log-likelihood, i.e., conditional on the studies being published. In the context of outcome reporting bias (ORB) adjustment methods, according to the framework developed by \citet{Copas2019}, the likelihood function takes into account studies for which we have both non-missing and missing outcome information. The studies have different log-likelihood contributions, depending on whether a study $i$ reports the outcome, i.e., $i \in \{\text{Rep} \}$, or the study $i$ does not report the outcome, i.e., $i \in \{ \text{Unrep} \}$. The full ORB-adjusted log-likelihood, where $K = K_{\text {Rep}}+ K_{\text {Unrep}}$ is the total number of studies, can be seen as

\begin{equation}
\label{lik.full}
\begin{aligned}
%L\left(\theta \right)&=\prod_{i=1}^n L_i\left(\theta \right) \quad \text { where } n=n_{\text {Rep }}+n_{\text {High }} \\
\ell_{\text{Adj}}^{\text{ORB}} & =\sum_{i=1}^K \ell \left(\theta \right)\\
& =\sum_{i \in \{ \operatorname{Rep}\} } \ell\left(\theta \right)+\sum_{i \in \{\operatorname{Unrep} \} } \ell\left(\theta \right) \\
& =\sum_{i \in \{\operatorname{Rep} \} } \log f \left(y_i \text{; } \theta \right)+\sum_{i \in \{\operatorname{Unrep} \} } \log f \left(y_i \text{; } \theta \right) \text{.}
\end{aligned}
\end{equation}

\bigskip

We can then adapt the formulation of equation \eqref{PB.selection11} for ORB, by considering, for reported studies $ \{i \in \text{Rep} \}$, the probability $w(y_i)$ of a study reporting an outcome, instead of the probability of a study being published. The following thus holds:

\bigskip

\begin{equation}
\label{ith.rep}
\begin{aligned}
f\left(y_i \text{; } \theta \mid  \, i \in \{\text{Rep}\} \right) = \frac{f\left(y_i \text{; } \theta \right) \cdot w(y_i)}{\int_{-\infty}^{\infty} f\left(y \text{; } \theta \right) \cdot w(y) d y} \text{.}
\end{aligned}
\end{equation}

\bigskip

Similarly, for the unreported studies $i \in \{\text{Unrep} \}$, we can use the formulation \eqref{PB.selection11} and consider the probability $1-w(y_i)$ of a study not reporting an outcome. We hence obtain

\bigskip

\begin{equation}
\label{ith.high}
\begin{aligned}
f\left(y_i \text{; } \theta \mid \, i \in \{\text{Unrep} \} \right) = \frac{f \left(y_i \text{; } \theta \right) \cdot \left(1-w(y_i) \right)}{\int_{-\infty}^{\infty} f \left(y \text{; } \theta \right) \cdot \left(1-w(y) \right) d y} \text{.}
\end{aligned}
\end{equation}

\bigskip

Using \eqref{ith.rep} and \eqref{ith.high}, and solving for $f(y_i \text{; } \theta)$, we can re-write the ORB-adjusted log-likelihood \eqref{lik.full} as

\bigskip

\begin{equation}
\label{lik.rewritten}
\begin{aligned}
\ell_{\text{Adj}}^{\text{ORB}} \left(\theta \right) & = \sum_{i \in \{\operatorname{Rep} \} } \log  f(y_i \text{; } \theta)\\
& - \sum_{i \in \{\operatorname{Rep} \} } \log \left[ \int_{-\infty}^{\infty} f\left(y \text{; } \theta \right) \cdot w(y) dy \right]\\
& + \sum_{i \in \{ \operatorname{Unrep} \}} \log \left[ \int_{-\infty}^{\infty} f(y \text{; } \theta) \cdot \left( 1 - w(y) \right) d y \right] \text{.}
\end{aligned}
\end{equation}

\bigskip

The log-likelihood \eqref{lik.rewritten} is the generic setting using a weight function for the probability of reporting, i.e., for $i \in \{\text{Rep} \}$, and a weight function for the probability of not reporting, i.e, for $i \in \{\text{Unrep} \}$. In the \citet{Copas2019} model formulation, specific assumptions were made regarding the missing data mechanism, which result in a simplification of \eqref{lik.rewritten}. For the reported outcomes, \citet{Copas2019} implicitly do not assume any selection process, i.e., $w(y_i)=1$ when $i \in \{\text{Rep} \}$. This means that no weight function representing the reporting probability based on the $p$-value is associated with the reported observations. In light of this assumption, \eqref{lik.rewritten} can be further simplified to

\begin{equation}
\label{lik.rewritten.special.copas.case}
\begin{aligned}
\ell_{\text{Adj}}^{\text{ORB}} \left(\theta \right) & = \sum_{i \in \{\operatorname{Rep} \} } \log f(y_i \text{; } \theta)\\
&+ \sum_{i \in \{ \operatorname{Unrep} \}} \log \left[ \int_{-\infty}^{\infty} f(y \text{; } \theta) \cdot \left( 1 - w(y) \right) d y \right] \text{.}
\end{aligned}
\end{equation}

%\subsubsection{Copas et al. Selection Function}

This is thus the generic form for ORB-adjustment, which has different shapes depending on the selection function $w(y_i)$ used, representative of the missing not at random (MNAR) mechanism assumed for unreported study outcomes. Given the alignment of our ORB-adjustment with the PB framework of selection models, one can use similar selection functions which are typically found in the PB literature. 

\subsection{Selection Functions} \label{ourSELs}

We present a series of selection functions implemented in our methodology, defined as functions of the one-sided $p$-value, $p = \Phi(- y / \sigma)$, where $\alpha$ is the threshold for significance, e.g., $\alpha=0.05$, and the study index $i$ is omitted for ease of notation. As previously noted, given that the selection mechanism in the following section is defined as a function of the $p$-value, we use the notation $w(p)$ instead of $w(y)$. 

\subsubsection{Piecewise constant}

We choose a one-sided $p$-value to model the probability of selection, in alignment with selection models of beneficial outcomes in PB \citep{HedgesVev, reviewselection, selectionCont}. One of the simplest selection functions used for PB is:

%\bigskip

\begin{equation}
\begin{aligned}
\label{sel0}
w_A(p)= \begin{cases} 1 & \text{if } p \leq \alpha  \\
0 & \text{if } p > \alpha \text{,} \end{cases}
\end{aligned}
\end{equation}

%\bigskip

While this selection function can be found in the PB literature \citep{selection0, selection1, reviewselection}, we note that it is also the one implicitly used in the \citet{Copas2019} adjustment, although the authors do not explicitly frame the ORB-adjustment via a selection model framework. Of note, in \citet{Copas2019}, ORB-adjustment is applied by including only the unreported study outcomes classified at HR of bias by the ORBIT classification system. They thus omit the unreported study outcomes classified at LR of bias and regard them as missing at random. Furthermore, the authors use the two-sided $p$-value $p = 2 \cdot (1 - \Phi(\abs{\frac{y}{\sigma}}))$ instead of the one-sided one proposed in this work. We deem a one-sided $p$-value to be more appropriate to model the underlying missing data mechanism for a beneficial effect of treatment, as it would be unlikely for significant outcomes, but in the wrong direction, to be reported \citep{HedgesVev, reviewselection, selectionCont}.

Using the log-likelihood \eqref{lik.rewritten.special.copas.case} and the selection function \eqref{sel0} for a subset of the unreported studies, i.e., those classified as HR of bias, along with a two-sided $p$-value instead of a one-sided one, we can easily see how we obtain the simplified ORB-adjusted log-likelihood presented for the random effects model in \citet{Copas2019}, namely:

\bigskip

\begin{equation*}
\label{w.copas.der}
\begin{aligned}
& \ell_{\text{Adj}}^{\text{ORB}}(\theta) = \sum_{i \in \{\text{Rep} \} } \log f(y_i \text{; } \theta)\\
& + \sum_{i \in \{ \text{HR} \}} \log \left[\int_{-\infty}^{+\infty} f(y \text{; } \theta) (1-w(y)) dy\right] \\
%& = \sum_{i \in \text{Rep}} \log f_i(y_i; \theta) + \sum_{i \in \text{HR}} \log \left[\int_{-z_{\alpha} \sigma_i}^{+z_{\alpha} \sigma_i} f_i(y ; \theta) \cdot 1 \cdot dy\right] \\
%& = \sum_{i \in \text{Rep}} \log f_i(y_i; \theta) + \sum_{i \in \text{HR}} \log \left[F_i(z_{\alpha} \sigma_i; \theta)-F_i(-z_{\alpha} \sigma_i; \theta)\right] \\
%& = \sum_{i \in \text{Rep}} \log f_i(y_i; \mu, \tau^2) + \sum_{i \in \text{HR}} \log \left[\Phi\left(\frac{z_{\alpha} \sigma_i - \mu}{\sqrt{\sigma_{i}^2 + \tau^2}}\right)-\Phi\left(\frac{-z_{\alpha} \sigma_i-\mu}{\sqrt{\sigma_{i}^2 + \tau^2}}\right)\right]\\
& = -\frac{1}{2}\sum_{i \in \{\text{Rep} \}  }\left[\log(\sigma_{i}^2 + \tau^{2}) + \frac{(y_{i}-\mu)^2}{\sigma_{i}^2+\tau^2}\right]\\
& + \sum_{i \in \{\text{HR}\}} \log \left[\Phi\left(\frac{z_{\alpha} \sigma_i - \mu}{\sqrt{\sigma_{i}^2 + \tau^2}}\right)-\Phi\left(\frac{-z_{\alpha} \sigma_i-\mu}{\sqrt{\sigma_{i}^2 + \tau^2}}\right)\right] \text{.}
\end{aligned}
\end{equation*}

\bigskip

\subsubsection{Constant-decreasing}

The selection function \eqref{sel0} results in a simple shape of the ORB-adjusted log-likelihood; however, the underlying assumption regarding the missing data mechanism is somewhat strict, and extensions which relax its assumption are commonly found in the PB literature \citep{selection1, reviewselection}. One example is the function $w_B(p \text{; } \beta)$ with tuning parameter $\beta >0$:

\begin{equation}
w_B(p \text{; } \beta) = 
\begin{cases}
1 & \text{if } p \leq \alpha \\
\frac{p^{-\beta}}{\alpha^{-\beta}} & \text{if } p > \alpha
\end{cases}
\label{special1}
\end{equation}

The idea of this selection function in the context of PB is that the associated probability of publishing, which weighs observations, is greater than 0 for non-significant outcomes. Specifically, when applied to ORB, for the unreported study outcomes which were originally non-significant, the underlying probability of reporting is a decreasing function of the $p$-value, while significant study outcomes have an associated probability of reporting equal to 1.

\subsubsection{Decreasing-constant}

In the context of ORB we further propose a different selection function, $w_C(p \text{; } \gamma)$ with tuning parameter $\gamma > 0$ presented in \eqref{special2}, for which the rationale is inverted compared to $w_B(p \text{; } \beta)$ in \eqref{special1}. With selection function $w_C(p \text{; } \gamma)$ we assume that non-significant study outcomes have an associated probability of reporting equal to 0, while significant study outcomes have an associated probability of reporting which is a decreasing function of the $p$-value. This can be motivated by scenarios where ORB results from prioritizing more impactful or clinically relevant findings in a published study \citep{moreORBreasons, ORBreasons}, leading to only highly significant outcomes being reported. This could be interpreted as a lower threshold for not reporting compared to PB, and thus a higher level of bias. At the same time, given that the selection function allows for significant unreported outcomes, it can also account for settings in which outcomes are missing because they were deemed less relevant, resulting in a pattern of missing data closer to missing completely at random (MCAR) and less bias \citep{mythesis}. Understanding the exact cause of unreporting can be challenging, and information on the strength of evidence for other outcomes in the meta-analysis could help clarify the likely cause of unreporting.

\bigskip

\begin{equation}
w_C(p \text{; } \gamma) = 
\begin{cases}
1 - \frac{p^{\gamma}}{\alpha^{\gamma}} & \text{if } p \leq \alpha \\
0 & \text{if } p > \alpha
\end{cases}
\label{special2}
\end{equation}

\bigskip

\subsubsection{Piecewise decreasing}

Based on the selection functions $w_B(p \text{; } \beta)$ in \eqref{special1} and $w_C(p \text{; } \gamma)$ in \eqref{special2} we  envisage a combination of these by using e.g., selection function $w_D(p \text{; } \beta, \gamma)$ in \eqref{special3}. In this case, one can flexibly specify both $\gamma$ and $\beta$ parameters, as well as the probability of reporting assumed for a study outcome at the significance threshold $\alpha$, which we note $\omega_{\alpha}$. In the case of \eqref{special1}, $\omega_{\alpha}$ was implicitly 1 and in case of \eqref{special2} this was set to 0. Here, we set $\omega_{\alpha}=0.5$, as a middle value between \eqref{special1} and \eqref{special2}. The selection function $w_D(p \text{; } \beta, \gamma)$ has the potential of being used to conduct extensive sensitivity analyses when adjusting for ORB.

\bigskip

\begin{equation}
w_D(p \text{; } \beta, \gamma) = 
\begin{cases}
1 - (1-\omega_{\alpha})\left(\frac{p^{\gamma}}{\alpha^{\gamma}}\right) & \text{if } p \leq \alpha \\
\omega_{\alpha}\left(\frac{p^{-\beta}}{\alpha^{-\beta}}\right) & \text{if } p > \alpha
\end{cases}
\label{special3}
\end{equation}

\bigskip

\begin{figure*}[!hbt]
\centering
\caption{Possible Selection Functions for ORB-adjustment. Function $w_A(p)$ from equation \eqref{sel0} in (a), function $w_B(p \text{; } \beta = 3)$ from equation \eqref{special1} in (b), function $w_C(p \text{; } \gamma = 3)$ from equation \eqref{special2} in (c), and functions $w_D(p \text{; } \beta = 1.5, \gamma = 7)$ and $w_D(p \text{; } \beta = 7, \gamma = 1.5)$  from equation \eqref{special3} shown in (d).\\[0.5em]}
%\hline

{\centering \includegraphics[width=\maxwidth]{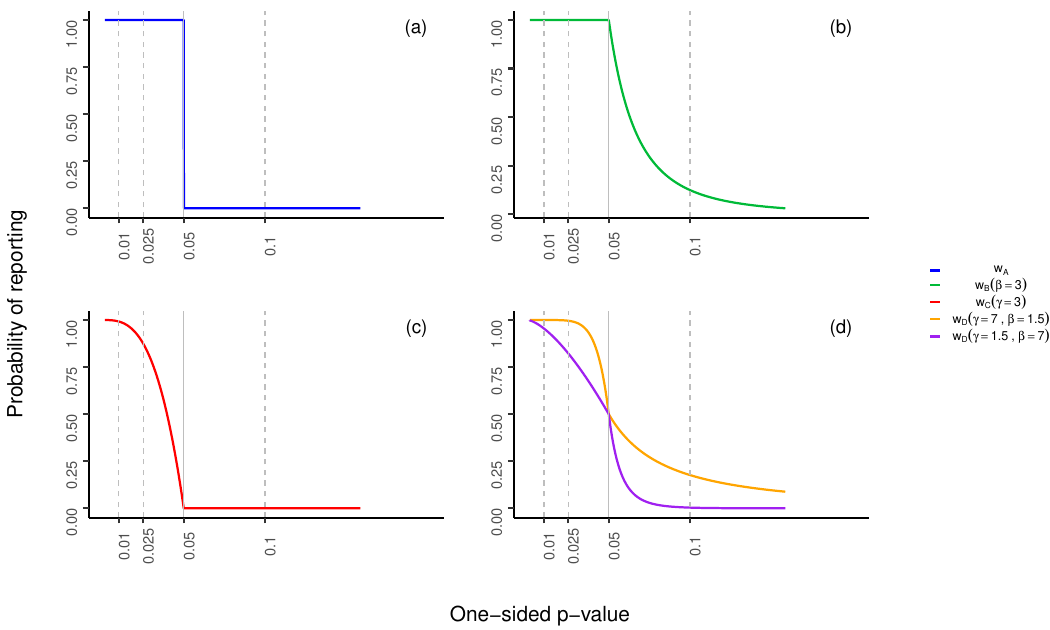} 

}

%\hline
\label{new.weight.fig}
\end{figure*}

The selection functions proposed above, namely $w_A(p)$ in \eqref{sel0}, $w_B(p \text{; } \beta, \gamma)$ in \eqref{special1}, $w_C(p \text{; } \gamma)$ in \eqref{special2} and $w_D(p \text{; } \beta, \gamma)$ in \eqref{special3} are plotted in Figure \ref{new.weight.fig} for some example values of the $\gamma > 0$ and $\beta > 0$ parameters. Further rationale for the parameter choices are discussed in the simulation study protocol, available in the \href{https://osf.io/ancdu/}{OSF project repository}.

\bigskip

With the selection model framework for ORB-adjustment presented in this work, one is thus able to include a likelihood contribution from unreported study outcomes, by specifying the desired missing data assumption via a selection function, representative of the assumed probability of reporting. This framework enables the joint estimation, via maximum likelihood (ML), of the ORB-adjusted parameters of interest in the model, in our case treatment effect, as well as the heterogeneity variance.

\subsection{Imputation of Missing Variances} \label{impvar}

When utilizing any of the selection functions presented in the ORB-adjusted log-likelihood \eqref{lik.rewritten.special.copas.case}, we require knowledge of the standard error of the unreported study outcome, which is generally missing. This value hence needs to be imputed; we follow the methodology of \citet{Copas2019}, \citet{Kirkham2012}, and \citet{Bay} and impute the missing standard error of an unreported study $i$ as

\begin{equation}
\label{k1}
\sigma_{i}^2 = \frac{1}{\hat{k} n_i} \text{,}
\end{equation}

%\bigskip

where $n_i$ is the sample size of study $i$ and

\begin{equation}
\label{k2}
\hat{k} = \frac{\sum_{i \in \{\text{Rep} \} } \sigma_{i}^{-2}}{\sum_{i \in \{\text{Rep} \}} n_i} \text{.}
\end{equation}

\bigskip

The rationale behind this imputation is grounded in the relationship between effect size variances and sample size. For commonly used measures such as the log risk ratio (log RR), the variance is inversely proportional to the sample size. However, this relationship is also influenced by the design of the trial, e.g., the balance between arms and the event rates, captured by the trial-specific design factor \(k_i = \sigma_i^{-2}/n_i\). While this factor varies across trials, it is assumed that the average design factor for the unreported studies is similar to that of the reported studies. Hence, we estimate this average using \eqref{k2} and apply it in \eqref{k1} to impute missing variances \citep{Copas2019, Copas2014}. This imputation strategy is particularly suitable in our setting because the total sample size \(n_i\) is typically known even when the outcome is not reported, due to the nature of ORB compared to, e.g., PB, where the whole study is missing from the literature.

\bigskip

%With the selection model framework for ORB adjustment presented in this work, one is thus able to include a likelihood contribution from unreported study outcomes, by specifying the desired missing data assumption via a selection function, representative of the assumed probability of reporting. This framework enables the joint estimation, via maximum likelihood (ML), of the ORB-adjusted parameters of interest in the model, in our case treatment effect, as well as the heterogeneity variance.

%\twocolumn[
%\begin{@twocolumnfalse}
\section{Simulation Study} \label{simstudy}
%\vspace{-\baselineskip} % Adjusts space after the section header
It is of interest to assess the extent to which outcome reporting bias (ORB) negatively impacts meta-analytic findings and the extent to which the ORB-adjustment methodology presented in the previous section of this work is effective in reducing bias. Our primary interest lies in bias detection and mitigation for treatment effect estimation under different meta-analysis settings, e.g., varying levels of heterogeneity and meta-analysis study sizes. A secondary interest of the investigation is the possible impact of ORB on heterogeneity variance estimation. In pursuit of these objectives, we conduct a simulation study wherein we first simulate a random effects meta-analysis of a single beneficial outcome and subsequently mimic selective reporting by removing some observed treatment effects and standard errors from the meta-analysis dataset based on the strength and/or direction of the results, favoring the reporting of studies with small $p$-values. The resulting data-generating mechanism (DGM) thus follows a missing not at random (MNAR) pattern, as reporting depends on statistical significance. We then utilize different estimation methods for the parameters of interest and assess the performance of the methods using performance measures on a large number of simulations. 

The details of the simulation study can be found in the simulation study protocol (already available in the \href{https://osf.io/ancdu/}{OSF project repository}) and are summarized in the following setting description section.

\subsection{Setting} \label{3.1}

The first step of the simulation process consists of simulating random effects meta-analysis datasets in the presence of ORB. We first simulate a random effects meta-analysis study comprising $K$ studies, each with treatment and control arms of equal sizes $n_i = n = 50$, and reported treatment effects $y_i$ with standard errors $\sigma_i$. We first obtain the study-specific true treatment effects $\theta$ from

\begin{equation*}\label{eq:random.eff2}
\theta_i \sim \mathcal{N}(\mu, \tau^2) \text{,}
\end{equation*}

where $\mu$ is the overall treatment effect and $\tau^2$ is the between-study heterogeneity variance. The observed treatment effects $y_i$ are then given by 

\begin{equation*}\label{eq:random.eff1}
y_i \sim \mathcal{N}(\theta_i, \sigma^2) \text{,}
\end{equation*}

where $\sigma^2= 2/n$, while the standard errors are generated from a scaled $\chi^2$ distribution

\begin{equation*}
\label{sigma.sq.sim}
\sigma_i^2 \sim \frac{\chi^2_{2n_i - 2}}{(n_i -1)n_i} \text{.}
\end{equation*}

These values are generated independently for each study, assuming no correlation between studies. We then simulate ORB by selectively excluding certain studies from the meta-analysis based on the direction and significance of treatment effects. The ORB simulation process involves removing study outcomes with a probability of reporting determined by a decreasing function of the one-sided $p$-value, i.e., $p_i = \Phi(- y_i / \sigma_i)$. The function \eqref{simORB} used to simulate ORB is taken from simulation studies on PB, for consistency with our selection model approach, typical of PB settings. We simulate under two ORB settings, i.e., $\gamma=1.5$ typical in PB simulation studies \citep{selection2, Begg, BeggUse1, selectionCont} and $\gamma=0.5$, resulting in a steeper decreasing function of the $p$-value. 

\begin{equation}
P( i \in \{\text{Rep}\})= e^{-4 \cdot p_i^{\gamma}} \text{.}
\label{simORB}
\end{equation}

Each meta-analysis dataset hence results in $K$ or fewer of the original study outcomes. If for some meta-analysis dataset less than two study outcomes are reported, the simulation is repeated until at least two reported study outcomes are obtained \citep{selection2, Begg, BeggUse1, FernandezSim}. The ORB-affected meta-analysis datasets are generated under different settings; we vary the number of studies in the meta-analysis, $K \in \{5, 15, 30 \}$, the amount of between-study heterogeneity $\text{I}^2 \in \{0 \%, 25 \%, 50 \%, 75 \%, 90 \% \}$ and the true underlying treatment effect $\mu \in \{0, 0.2, 0.4, 0.6, 0.8 \}$, based on simulation studies found in the literature \citep{MorenoSim, IntHout2014, FernandezSim}.

It is important to note that while our simulation uses a known missing data mechanism, i.e., selective reporting based on a continuous function of the one-sided $p$-value, no information about the cause of missingness is assumed to be available once the data are generated. This reflects the practical setting in which some outcomes are unreported, but their classification into, e.g., high risk (HR) or low risk (LR) of bias as per the ORBIT methodology is unknown. This differs from the simulation study in our previous work \citep{mythesis}, where unreported outcomes were simulated mimicking the ORBIT classification into HR and LR of bias, and this information was assumed to be available prior to the ORB-adjustment. We deem our current way of simulating ORB, as a continuous function of the $p$-value, more realistic, as it is consistent with the selection model literature, less arbitrary than mimicking HR/LR of bias missingness, and, overall, more aligned with the goal of avoiding the need to for an ORBIT classification for ORB-adjustment. As a result however, the \citet{Copas2019} method, which requires the ORBIT classification and adjusts only for outcomes deemed at HR of bias, is not applicable in this setting and thus not directly comparable to our methodology.

After having simulated ORB, hence resulting in some treatment effects and standard errors unreported, we use maximum likelihood (ML) estimation to obtain point estimates of the treatment effect $\mu$ and the heterogeneity variance $\tau^2$, along with profile likelihood (PL) confidence intervals (CI) \citep{tauCI, Copas2019, likelihood}. The ML estimate and PL CI for $\mu$ and $\tau^2$ are obtained using different log-likelihoods, depending on the information and/or missing data mechanism assumed, leading to i) naive, ii) complete data, and iii) ORB-adjusted estimation methods. We further differentiate various ORB-adjusted estimates based on the selection function assumed for the probability of reporting.

The naive log-likelihood (i) includes the contribution only from reported study outcomes and disregards the unreported ones. The naive estimate serves as a baseline for comparison of the ORB-adjustment methodologies and quantifies the negative impact of ORB when the latter is not accounted for \citep{reviewselection, Copas2019}. The complete data log-likelihood (ii) uses all studies in the meta-analysis before ORB is simulated, and is a proxy for the true treatment effect if there were no ORB. The various ORB-adjusted estimates (iii) are obtained by maximizing the ORB-adjusted log-likelihood \eqref{lik.rewritten.special.copas.case} using the different selection functions:  $w_A(p)$ from \eqref{sel0}, $w_B(p \text{; } \beta=3)$ from \eqref{special1}, $w_C(p \text{; } \gamma=3)$ from \eqref{special2}, and $w_D(p \text{; } \beta=1.5, \gamma=7)$, $w_D(p \text{; } \beta=7, \gamma=1.5)$ from \eqref{special3}, as well as the selection function \eqref{simORB} used to simulate ORB, so as to include the correct model specification in the adjustment. Since the latter function used to simulate ORB can be viewed as a selection function, defining a specific missing data mechanism, we note it as $w_{DGM}(y)$ and utilize it in the ORB-adjusted log-likelihood \eqref{lik.rewritten.special.copas.case}. The parameters of the selection functions, i.e., $\beta$ or $\gamma$ used in the adjustment correspond to those illustrated in Figure \ref{new.weight.fig}.\\
\textcolor{white}{T} For each parameter setting, the simulation process is repeated $\text{N}_{\text{sim}} = 3200$ times; the simulation size $\text{N}_{\text{sim}}$ is calculated based on the expected variance of the unknown parameter estimate \citep{IntHout2014, sim} and a desired Monte Carlo Standard Error (MCSE) of 0.005 from \citet{IntHout2014, sim}. The performance measures recorded for the unknown parameter are bias, empirical standard error (ESE), mean squared error (MSE), coverage and power, along with the MCSEs of each \citep{IntHout2014, sim}.

To provide a methodological benchmark, we also simulate an alternative missing data mechanism under MCAR. Specifically, for each ORB setting in the above MNAR simualation process with $\gamma = 1.5$, we generate a corresponding MCAR scenario by removing the same number of studies as were dropped due to ORB, but at random and independently of their $p$-values. The estimation methods remain unchanged, with the naive estimation now aligned with the DGM under MCAR, while all ORB-adjusted methods are misspecified, as they assume a MNAR under MCAR. This methodological variation enables an assessment of the robustness of our main simulation process and ORB-adjustment approaches when the missingness is not due to ORB. A synthesis of the results is presented in Section \ref{mcar}, with additional details available in the supplementary material.

\subsection{Results} \label{3.2}

%%%%%% bias with naive estimation
\subsubsection{Bias in Naive Estimation}
The results demonstrate a significant bias in the estimation of the treatment effect when using naive methods that do not account for ORB, as shown in Figures \ref{res1} and \ref{res2}. The bias decreases as the true treatment effect size $\mu$ increases, which aligns with existing literature \citep{Copas2019, Bay, dutch} and prior exploratory analysis \citep{mythesis}; as treatment effect increases, results are more likely to be significant and are thus less prone to ORB. Study size variations ($K=5, 15, 30$) do not significantly affect the bias, while heterogeneity has a substantial impact. High heterogeneity settings, particularly with $I^2=90$, exhibit larger biases, reinforcing findings from previous work \citep{mythesis}. The effect of heterogeneity on ORB is interesting and novel compared to \citet{Copas2019}, who focused primarily on a fixed effect meta-analysis framework. The patterns observed for naive estimation are consistent across both ORB simulation processes, i.e., for $\gamma=1.5$ and $\gamma=0.5$ in the DGM function \eqref{simORB}.

%%%% bias reduction with DGM and other selection functions
%%%% NB for k=5 not good
\subsubsection{Bias Reduction with ORB-adjustment}
\begin{figure*}[!hbt]
\centering
\caption{Bias in the estimation of the treatment effect $\mu$ for ORB simulated according to DGM function \eqref{simORB} with $\gamma=1.5$, using different estimation methods, i.e., naive or ORB-adjusted according to the various selection functions indicated in the legend. The bias is shown for varying meta-analysis study sizes, heterogeneity levels, and increasing true treatment effect on the x-axis of each plot shown.\\[0.5em]}
%\hline

{\centering \includegraphics[width=\maxwidth]{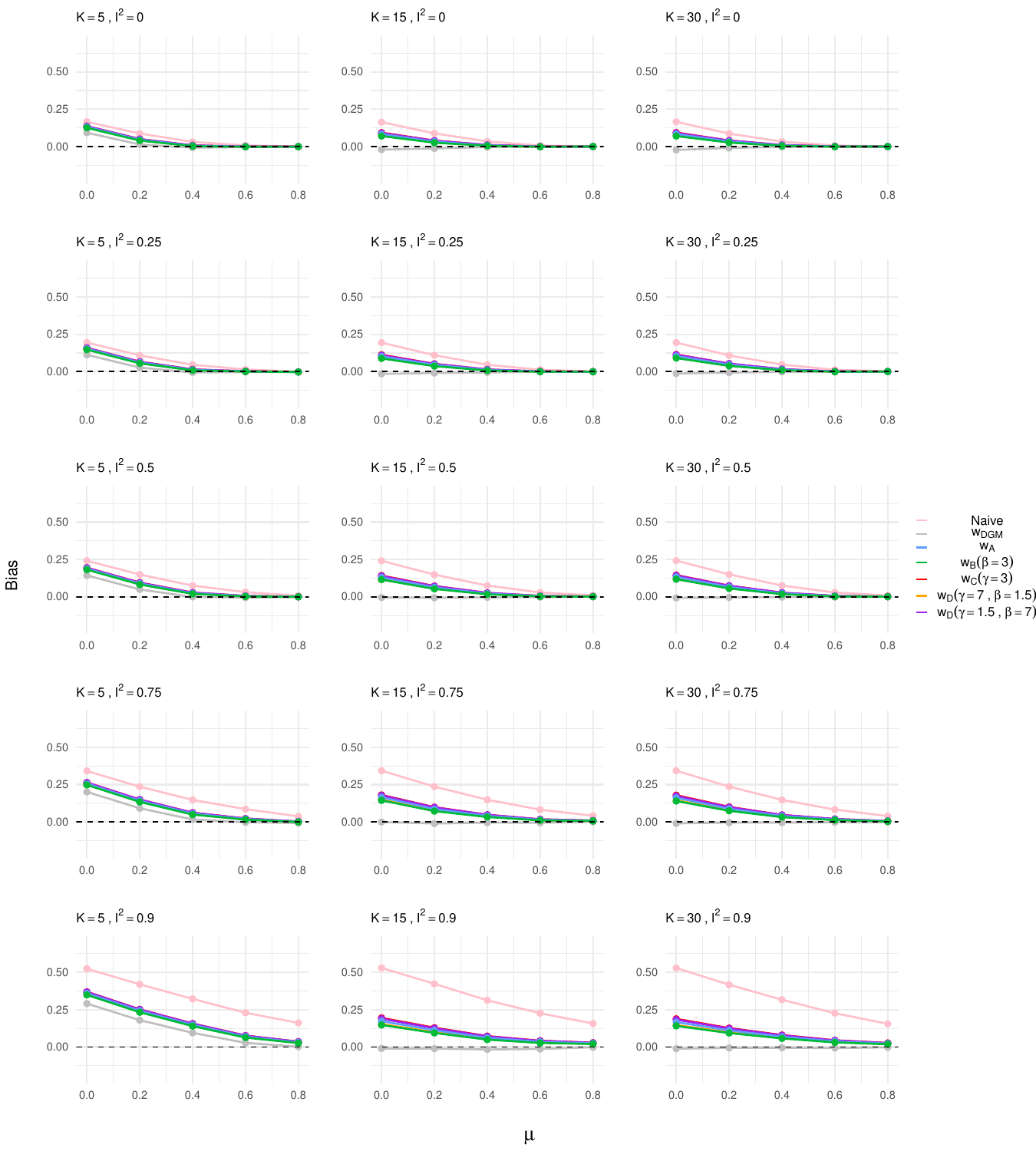} 

}

%\hline
\label{res1}
\end{figure*}

\begin{figure*}[!hbt]
\centering
\caption{Bias in the estimation of the treatment effect $\mu$ for ORB simulated according to DGM function \eqref{simORB} with $\gamma=0.5$, using different estimation methods, i.e., naive or ORB-adjusted according to the various selection functions indicated in the legend. The bias is shown for varying meta-analysis study sizes, heterogeneity levels, and increasing true treatment effect on the x-axis of each plot shown.\\[0.5em]}
%\hline

{\centering \includegraphics[width=\maxwidth]{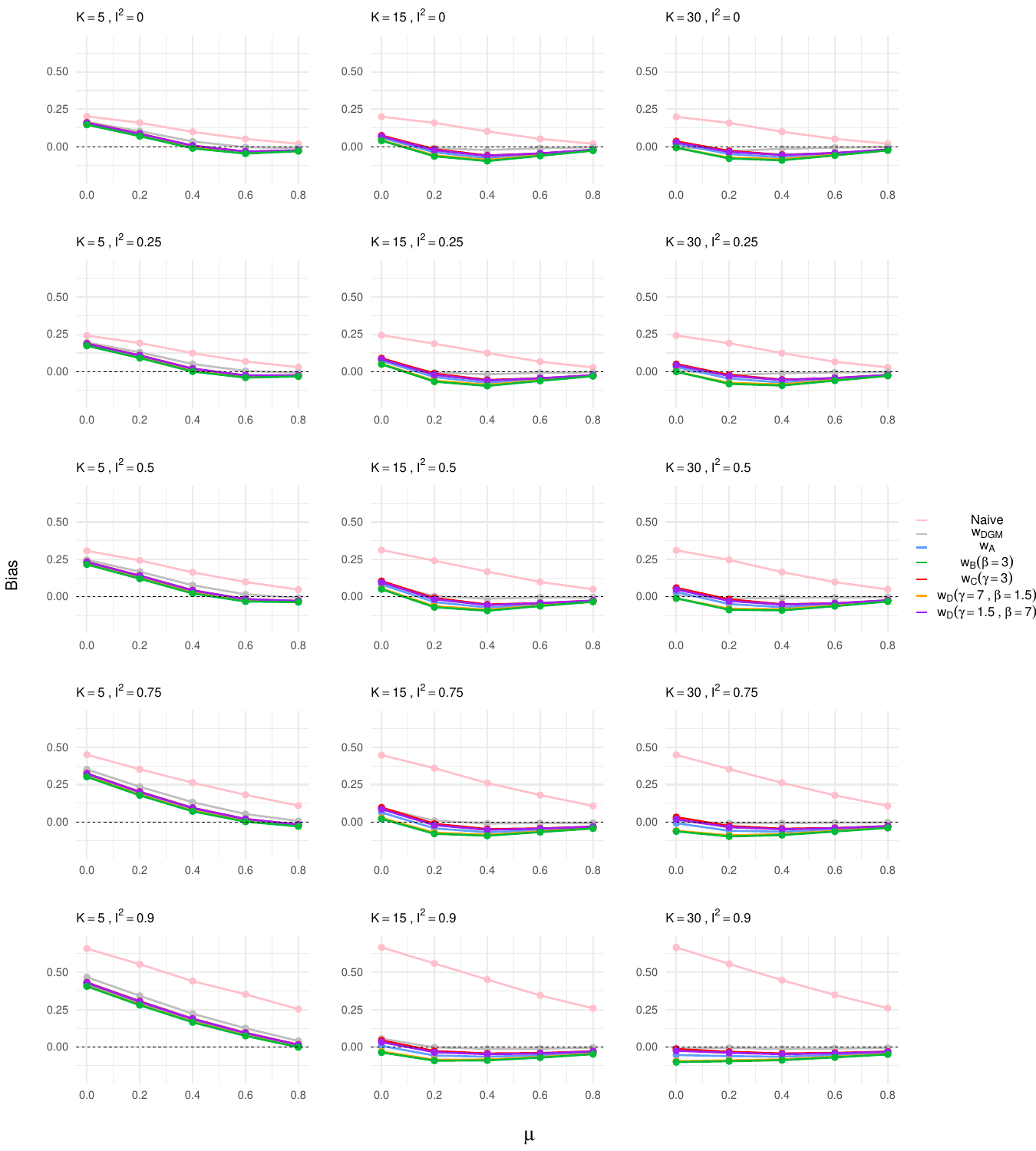} 

}

%\hline
\label{res2}
\end{figure*}

When applying the ORB-adjustment framework using selection functions, we first note that the effectiveness in bias reduction varies by meta-analysis study size. For $K=15$ and $K=30$, the bias obtained with naive estimation is eliminated when the selection function matches the ORB DGM, i.e., when using the correctly specified selection function $w_{DGM}$ in Figures \ref{res1} and \ref{res2}. Different selection functions ($w_A, w_B, w_C, w_D$) show varying degrees of bias reduction. For the DGM with $\gamma=1.5$, shown in Figure \ref{res1}, the ORB-adjusted estimates shift the bias towards the null but do not fully eliminate it unless the exact DGM function is used. The ORB-adjusted estimate using selection function $w_B$ performs slightly better than $w_A$, and $w_C$ performs the least well. However, it is important to note that the different functions ($w_A, w_B, w_C, w_D$) overall achieved very similar results, particularly in low heterogeneity settings. Similar patterns are observed in the DGM with $\gamma=0.5$ setting shown in Figure \ref{res2}, with selection function $w_B$ being the least strict and $w_C$ the most strict in bias reduction. The ORB-adjustment here tends to reduce the treatment effect size excessively, indicating potential overcorrection due to the steep $p$-value dependence in the ORB DGM, resulting in some unreported studies with significant $p$-values. 

For the small meta-analysis size, $K=5$, the ORB-adjustment reduces the bias but does not eliminate it, even with the correctly specified model. This finding holds in general for both ORB DGM settings; notably for the DGM with $\gamma=1.5$, the correctly specified selection function shows the least bias, while for the DGM with $\gamma=0.5$, it shows the most bias, although, overall, the ORB-adjusted estimates are similar. Based on these observations, we thus recommend to use ORB-adjustment with caution when only few studies are present in a meta-analysis affected by ORB.

\subsubsection{Other Performance Measures}

\begin{figure*}[!hbt]
\centering
\caption{Coverage in the estimation of the treatment effect $\mu$ for ORB simulated according to DGM function \eqref{simORB} with $\gamma=1.5$, using different estimation methods, i.e., naive or ORB-adjusted according to the various selection functions indicated in the legend. The coverage is shown for varying meta-analysis study sizes, heterogeneity levels, and increasing true treatment effect on the x-axis of each plot shown.\\[0.5em]}
%\hline

{\centering \includegraphics[width=\maxwidth]{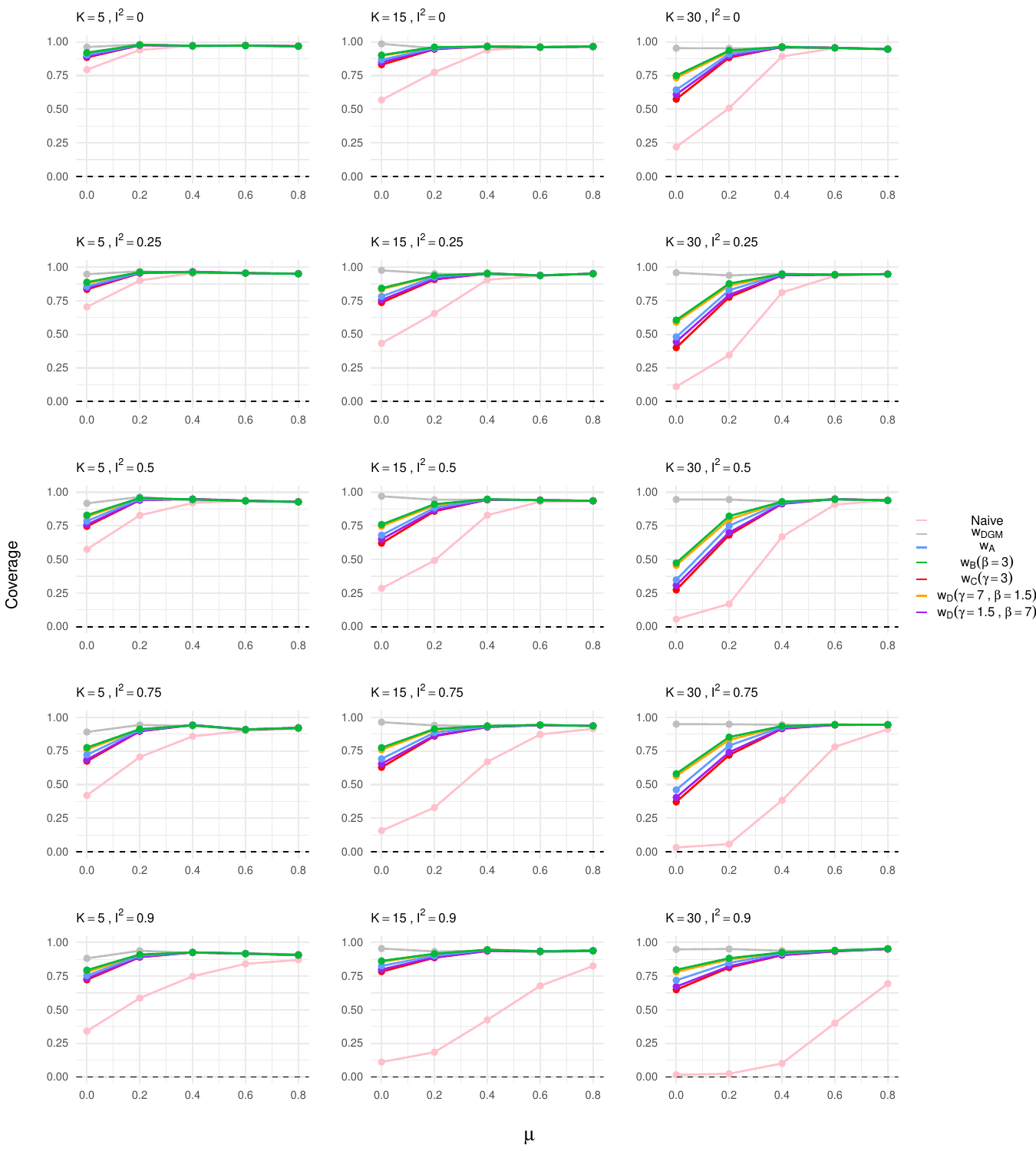} 

}

%\hline
\label{Cov1}
\end{figure*}

Beyond bias, other performance measures such as coverage, mean squared error (MSE), power, and empirical standard error (ESE) were evaluated. Coverage, shown for the ORB DGM $\gamma=1.5$ in Figure \ref{Cov1}, can be substantially low for naive estimation. Overall, the coverage in naive estimation decreases as heterogeneity increases. For small treatment effect sizes (e.g., $\mu=0$), coverage is higher for small meta-analysis sizes ($K=5$) and decreases as the meta-analysis size increases. This can be explained by larger CI for the $K=5$ setting, which in turn cover the true underlying value. The ORB-adjusted estimates show higher coverage: the correct DGM selection function has the highest one, while others exhibit slightly lower coverage, especially when $\mu$ is small. 

Other performance measures confirm the findings observed for naive and ORB-adjusted estimates, e.g., the MSE of the naive estimate of the treatment effect is substantially reduced in high heterogeneity settings for all ORB-adjusted estimates. Furthermore, naive estimation results in severely inflated power, particularly in high heterogeneity settings and for large meta-analysis study sizes. ORB-adjusted estimates correct this inflation, with variations depending on the DGM and selection function used. The ESE of the naive estimate is generally consistent with expected SE calculations from the simulation study protocol. Naive estimates have slightly higher SE due to unreported study outcomes. For small meta-analysis sizes ($K=5$) and high heterogeneity ($I^2=90$), ORB-adjusted estimates have a similar ESE, which is lower than the naive estimate. For detailed results and further plots of the additional performance measures considered and briefly mentioned in this section, please refer to the supplementary material.

\subsubsection{Bias in Heterogeneity Variance}

\begin{figure*}[!hbt]
\centering
\caption{Bias in the estimation of the heterogeneity variance $\tau^2$ for ORB simulated according to DGM function \eqref{simORB} with $\gamma=1.5$, using different estimation methods, i.e., naive or ORB-adjusted according to the various selection functions indicated in the legend. The bias is shown for varying meta-analysis study sizes, true treatment effect values, and increasing heterogeneity on the x-axis of each plot shown.\\[0.5em]}
%\hline

{\centering \includegraphics[width=\maxwidth]{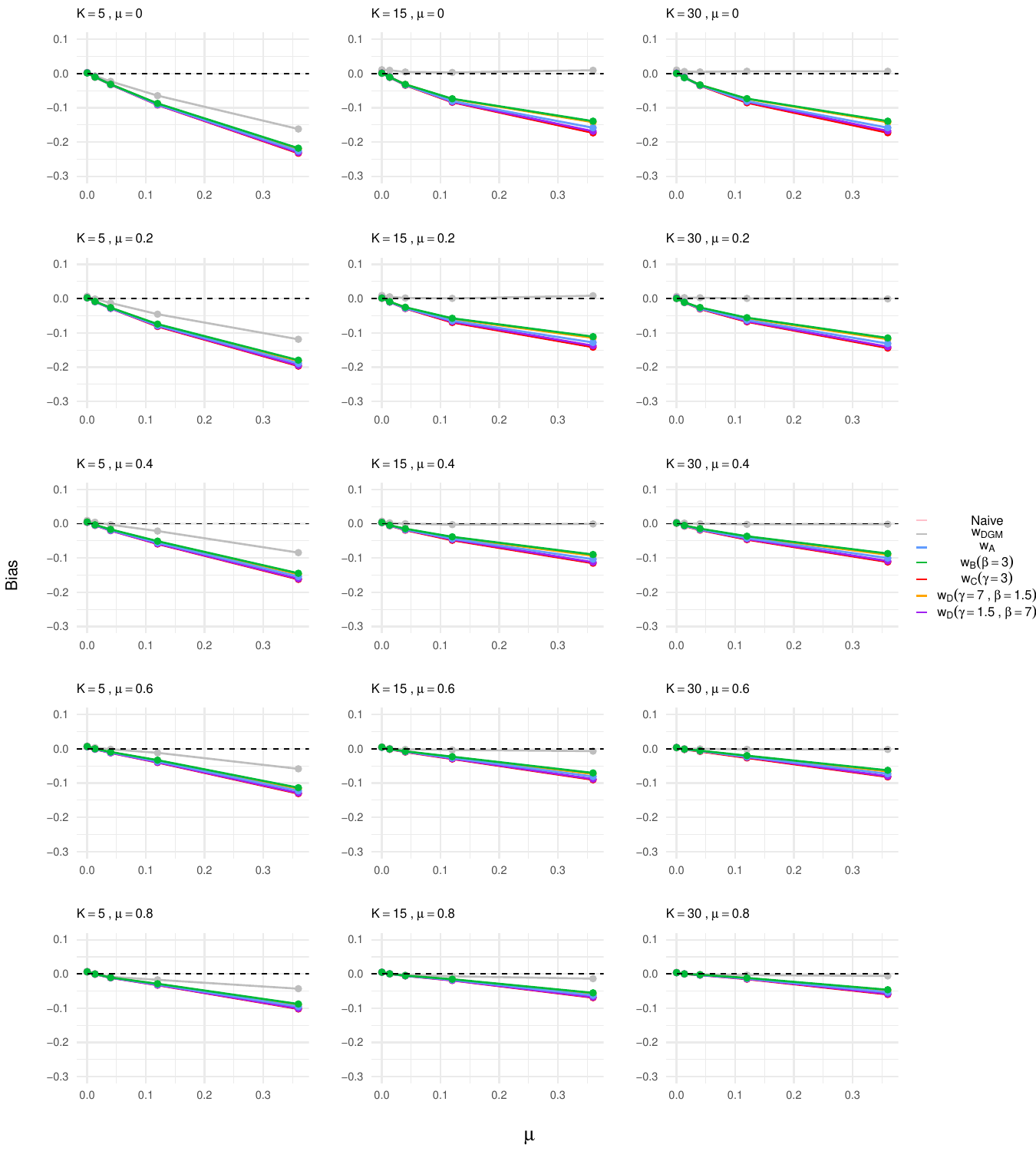} 

}

%\hline
\label{biastau1}
\end{figure*}

%%%% Some comments on tau^2, but only bias
Although the primary parameter of interest was the treatment effect $\mu$, we also investigated the bias in the estimation of the heterogeneity variance $\tau^2$ in the presence of ORB, as shown in Figure \ref{biastau1}, showcasing the results for the ORB DGM with $\gamma=1.5$. Heterogeneity is generally underestimated across most estimation methods, except for the ORB-adjusted method using the correctly specified model ($w_{DGM}$) for $K=15,30$. For the small meta-analysis setting $K=5$, the correctly specified model reduces the bias but does not fully eliminate, similarly to results observed for the main parameter of interest $\mu$. Of note, the estimation of $\tau^2$ is done with maximum likelihood (ML) estimation, which, overall, tends to underestimate the between study-heterogeneity \citep{REML2, REML}; a more comprehensive methodological approach to heterogeneity estimation in the presence of ORB should thus be conducted to solidify and confirm these findings. The plots of the heterogeneity variance for ORB setting $\gamma=0.5$ can be found in the supplementary material.

\subsubsection{Simulation under MCAR} \label{mcar}

Simulations under a MCAR mechanism (results presented in Figures $9$-$10$ of the supplementary material) showed that with few studies ($K = 5$) and near-null true treatment effects, the naive estimator was biased, reinforcing the caution advised in our primary simulation regarding ORB-adjustment in small meta-analyses with unreported study outcomes. In larger meta-analyses, the naive estimator was unbiased under MCAR, while ORB-adjusted methods, as expected due to model misspecification, underestimated the treatment effect, but remained reasonably robust, particularly in settings with low heterogeneity. These findings further support the use of ORB-adjustment methods when ORB is suspected and underscore the importance of conducting sensitivity analyses, given the inherent uncertainty of the true missing data mechanism.

%shown the bias in the estimation for the various meta-analysis sizes and for different true treatment effect values, we observe that as heterogeneity increases, the bias in its estimation for increases, i.e., in high heterogeneity settings, naive estimation underestimates the heterogeneity variance. For no heterogeneity or really low, this does not seem to be an issue. The ORB-adjusted estimate using the correct DGM mechanism appears to reduce this bias, while the other misspecified functions only partially reduce the bias, with no remarkable differences noted. We observe that this bias is more prominent for low values of the true underlying treatment effect, while for high values of $\mu$ the bias in the estimation of the heterogeneity is minimal, which can be attributed to the lack of ORB issue as there is less likelihood of unreporting.

\subsection{Application to Example Data} \label{3.3}

\begin{figure*}[!hbt]
\centering
\caption{Application of ORB-adjustment to the epilepsy example \citep{Copas2019, topiramate}, using the different selection functions shown in Figure \ref{new.weight.fig}. In addition, the naive estimate, without ORB-adjustment, is also given. The plot shows the risk ratio (RR) estimates and $95\%$ profile likelihood (PL) confidence intervals (CI).}
%\hline

{\centering \includegraphics[width=\maxwidth]{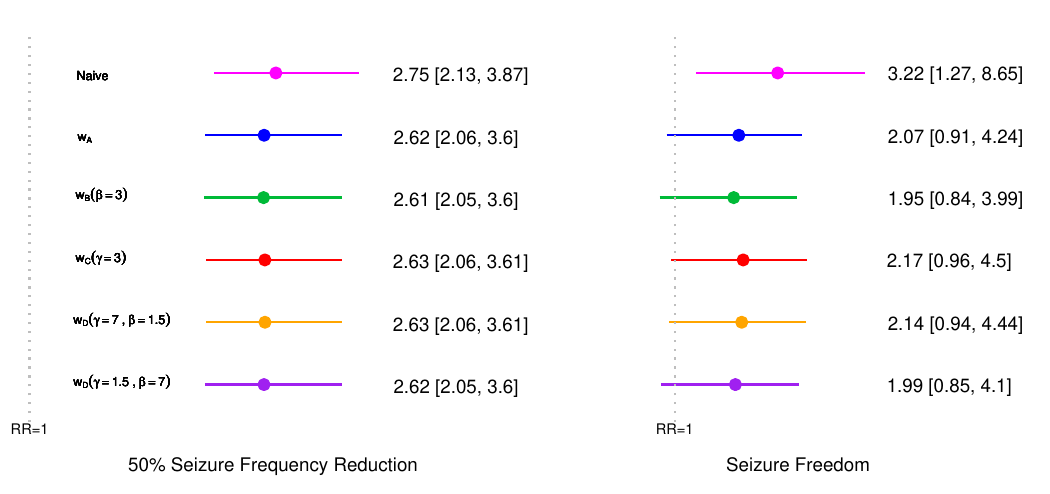} 

}

%\hline
\label{example}
\end{figure*}

\begin{figure*}[!hbt]
\centering
\caption{Application of ORB-adjustment sensitivity analysis to the epilepsy example \citep{Copas2019, topiramate}, using selection function \eqref{special1} in (a), selection function \eqref{special3} in (b), and selection function \eqref{special2} in (c).\\[0.5mm]}

{\centering \includegraphics[width=\maxwidth]{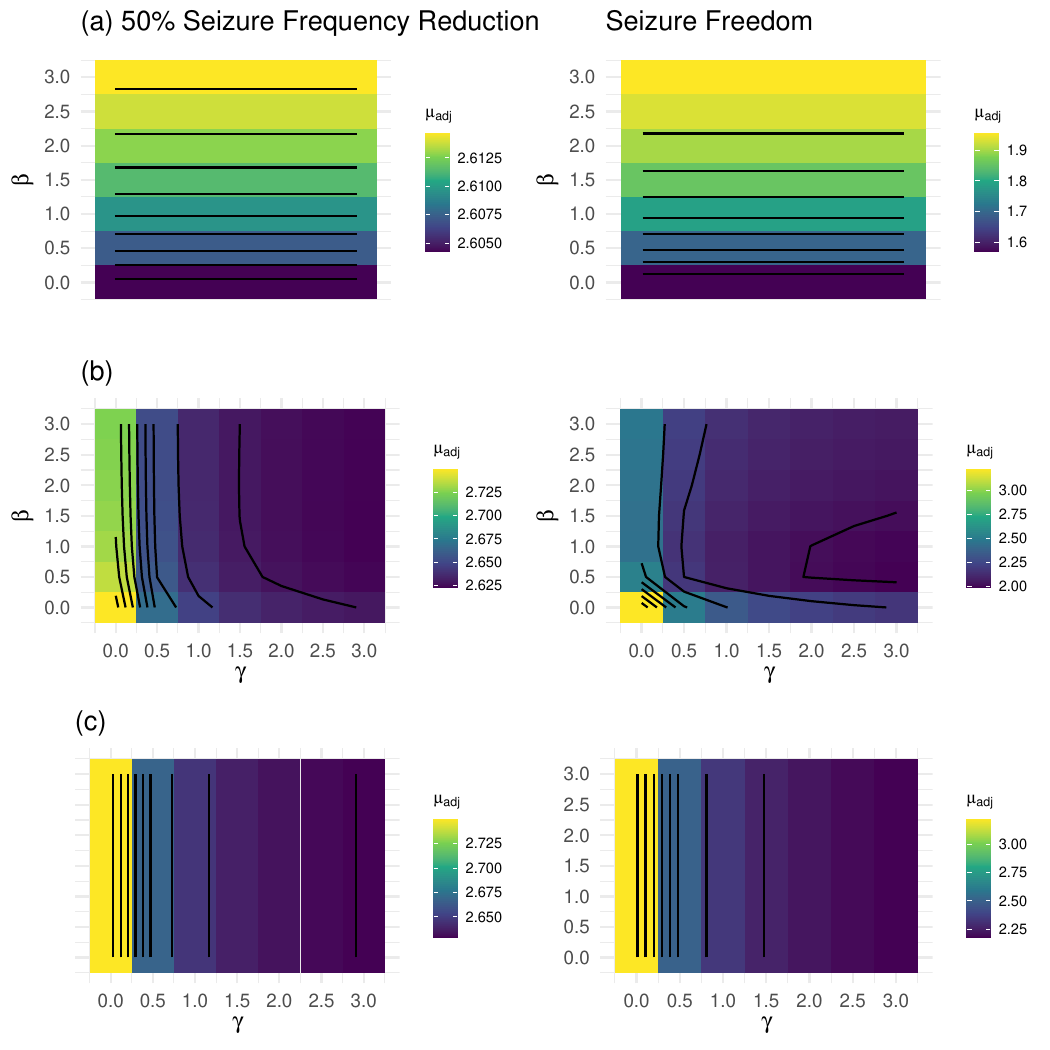} 

}

\label{exampleSENS}
\end{figure*}

We apply the ORB-adjustment framework to the epilepsy example from from \citet{topiramate} and \citet{Copas2019} introduced in Table \ref{tab}. We adjust for ORB in the two beneficial outcomes of the example meta-analysis, i.e.,  $50\%$ reduction in seizure frequency and seizure freedom using the selection functions proposed in this research and utilized in the simulation study. Figure \ref{example} shows the results of applying the same parameters used in the simulation study to our example data.  Figure \ref{exampleSENS} shows the results of an example sensitivity analysis for the ORB-adjusted estimates for a range of such parameters. %These selection functions utilize a one-sided $p$-value for significance with a threshold of $\alpha=0.05$, in contrast to \citet{Copas2019}, which used a two-sided $p$-value. Of note, while a one-sided threshold is used within the selection function to define the underlying missing data mechanism, two-sided significance is used to construct profile likelihood (PL) confidence intervals (CIs) for treatment effect estimation.

Figure \ref{example} presents the point estimates and $95\%$ CI for the RR of the treatment effect for the two $50\%$ reduction in seizure frequency and seizure freedom. For both outcomes, the naive RR estimate, i.e., that obtained with standard meta-analysis methods, thereby excluding any contribution from unreported study outcomes, shows a significant positive treatment effect compared to the control.

For the $50\%$ seizure frequency reduction outcome, the ORB-adjusted estimates are slightly shifted towards the null value and are consistent across different selection functions. Only a minor shift is expected since only one study does not report this outcome. However, for the seizure freedom outcome, with several studies not reporting it, the ORB-adjusted estimates show a substantial shift towards the null, even altering the significance of the results by causing the $95\%$ CI to overlap with 0. The differences between the ORB-adjusted estimates using various selection functions are more pronounced for the seizure freedom outcome.

The strictness of different ORB-adjustments varies depending on the assumptions of each selection function. The estimate obtained using selection function $w_B(\beta=3)$ from \eqref{special1} is more conservative than that obtained with $w_A$ in \eqref{sel0}. The selection function $w_A$ assumes a probability of unreporting of 1 for non-significant studies, regardless of the $p$-value magnitude, while $w_B(\beta=3)$ assumes a higher probability of unreporting for larger $p$-values, implying greater bias and thus stricter correction. Conversely, $w_C(\gamma=3)$ from \eqref{special2} is less conservative than $w_A$, as it assumes that some unreported outcomes may still be significant, indicating less bias and thus a less strict ORB-adjustment. The hybrid functions $w_D(\gamma=1.5,\beta=7)$ and $w_D(\gamma=7,\beta=1.5)$ combine these behaviors, and their estimates lie between those of $w_B$ and $w_C$.

Given that the true missing data mechanism is unknown, we recommend a sensitivity analysis approach to ORB-adjustment, using different selection functions (e.g., $w_B(\beta)$ from \eqref{special1}, $w_C(\gamma)$ from \eqref{special2} or $w_D(\gamma, \beta)$ from \eqref{special3}) and varying parameters $\beta$ and $\gamma$, which control how strongly the probability of unreporting depends on the $p$-value. Figure \ref{exampleSENS} showcases an example of a sensitivity analysis to ORB-adjustment using the three above-mentioned different selection functions and a range of $\beta$ and $\gamma$.

From the results in Figure \ref{exampleSENS} we observe that the $50\%$ seizure frequency reduction outcome, with only one unreported study, is minimally sensitive to parameter changes, while the seizure freedom outcome, with several missing results, exhibits greater variability. We note that smaller values of $\beta$ shift the selection mechanism $w_B(\beta)$ away from the less strict piecewise constant function $w_A$ (where all non-significant results have the same probability of unreporting of 1); therefore, smaller $\beta$ lead to stricter ORB-adjustments. On the other hand, as $\gamma$ increases, the function $w_C(\gamma)$ approaches the stricter behavior of $w_A$ (where all significant results have the same probability of unreporting 0); hence, for lower $\gamma$ values, we observe a less strict ORB-correction. The combined function $w_D(\gamma, \beta)$ reflects an interplay of both effects.

Overall, the parameter variations tested here ($\gamma=0$ to $3$, $\beta=0$ to $3$) do not drastically alter the results in Figure \ref{example} but reveal how selection function shape and strictness can impact the magnitude of adjustment. Nevertheless, since the true mechanism driving outcome reporting bias is unknown in practice, we strongly recommend using a sensitivity analysis approach with multiple plausible selection functions and parameter combinations. This enhances the robustness of conclusions and helps quantify the potential impact of ORB on meta-analytic estimates.

\section{Discussion} \label{disc}

This study addresses Outcome Reporting Bias (ORB), where the significance of study outcomes influences their reporting, leading to overestimation of beneficial treatment effects in meta-analyses of clinical trials. We approached ORB-adjustment through a selection model framework, a common method in publication bias (PB) literature. This methodology allows one to incorporate contributions from unreported study outcomes based on different assumed missing data mechanisms, specified via selection functions. Our proposed selection functions expand on existing methods, including those from previous works like \citet{Copas2019}, by being more flexible in the missing data mechanisms assumed, utilizing information from all identified unreported study outcomes, and jointly estimating both treatment effect and heterogeneity variance via a random effects model.

We conducted a simulation study on ORB and our ORB-adjustment methodology. The findings of our simulation study reveal several critical insights regarding the impact of ORB on the estimation of treatment effects and the efficacy of ORB-adjustment techniques. Naive estimation methods that do not account for ORB exhibit substantial bias, particularly in high heterogeneity settings, underscoring the importance of incorporating ORB-adjustments. 

Our results demonstrate that ORB-adjustment frameworks using selection functions can significantly reduce bias, although their effectiveness may vary with meta-analysis study size and the underlying method used to simulate ORB. For larger meta-analyses ($K= 15, 30$), correctly specified ORB-adjustment models effectively eliminate bias. Different misspecifications of the assumed missing data mechanism can be either slightly too lenient or slightly too strict, though their performance does not vary significantly. For smaller meta-analyses ($K=5$), we must be cautious as bias reduction is limited even with correctly specified models. 

Other measures of performance confirm these findings, demonstrating, e.g., substantial improvements in the coverage and power of the treatment effect estimates with ORB-adjustment. These findings highlight the necessity of using ORB-adjustment methods to achieve more accurate treatment effect estimates. Additionally, they suggest that heterogeneity estimation is impacted by ORB, warranting further attention to improve the robustness of meta-analyses in the presence of ORB. 

We applied our ORB-adjustment methodology to a real-world meta-analysis of epilepsy trials \citep{Copas2019, topiramate} affected by ORB and conducted an example sensitivity analysis approach under different strengths of ORB-adjustment. The ORB-adjusted estimates of the treatment effect were substantially shifted towards null values, compared to the naive estimate, i.e., the standard estimation not accounting for ORB. This shift was particularly significant in the presence of numerous unreported study outcomes. Varying the parameters in the sensitivity analysis did not strongly impact the ORB-adjusted estimates; however, we highlight that this was done on only one example data and for a limited set of parameters. Therefore, in practice we strongly recommend a sensitivity analysis approach to ORB-adjustment through selection models. 

The ORB-adjustment methodology via selection models proposed in this research is flexible and broadly applicable. Although promising results have been observed, several limitations exist and should be noted to promote future research in this field. Firstly, our framework operates on individual outcomes in meta-analyses, not accounting for correlations between outcomes. Future research could explore methods to incorporate the explicit modelling of the missing data mechanism through selection functions into the multivariate meta-analysis framework \citep{Liu2017, Hwang2018, Bay}, as well as continue addressing the limitations in the estimation of correlations in the presence of ORB \citep{Bay, Kirkham2012, Hwang2018, Liu2017}.

Another possible avenue of improvement in our current approach is the imputation of missing variances, as described in Section \ref{impvar} and done in previous works \citep{Copas2014, Copas2019, Bay, mythesis}. While this did not greatly impact our results due to equal study sizes in the simulation study setup, alternative ways to estimate the missing variances, e.g., via multiple imputation, could be considered \citep{var_imp, var_imp2}. Additionally, our ORB-adjustment methodology assumes normally distributed outcomes, which might not be precise for binary data \citep{Copas2014, Copas2019}, especially in cases of zero/low event numbers such as the epilepsy example from \citet{Copas2019}. Exploring a binomial likelihood for ORB-adjustment could be a potential avenue, as noted in \citet{mythesis}. In \citet{mythesis} we set-up the binomial likelihood contribution of reported studies, which can be extended to include a term from unreported studies with a specified probability of reporting.

We established that heterogeneity variance estimation is affected by ORB, and, at the same time, the true underlying heterogeneity influences the bias in the treatment effect estimate due to ORB. Therefore, considering heterogeneity in ORB and ORB-adjustments is of paramount importance. To address this, we focused on and conducted simulations using the random effects model, in contrast to \citet{Copas2019}, which concentrated on the fixed effects model. Maximum likelihood estimation (MLE) was used for estimating heterogeneity variance due to its connection to ORB-adjustment, i.e.,  the ORB-adjustment itself is defined via a likelihood function contribution. More sophisticated methods in the likelihood framework, such as restricted maximum likelihood (REML), could be considered \citep{REML, tauCI, REML2, mythesis}. An exploratory REML approach was proposed in previous work \citep{mythesis}, but a more robust derivation could be investigated. Obtaining accurate estimates of $\tau^2$ is crucial, and while challenging to intertwine it with ORB-adjustment outside the likelihood framework of joint estimation with $\mu$, novel methods could be investigated \citep{tauCI, REML, REML2}. Another potential area for future research is the effect of ORB on prediction intervals \citep{PI_coverage} and how ORB-adjustments impact them, as mentioned in previous work \citep{mythesis}.

One additional avenue for future research on ORB is the multiple imputation (MI) of missing study outcomes, which has also been used in publication bias (PB). In the context of PB, \citet{MI_PB_Carpenter2011} fit a model to the observed study data and impute missing studies using a missing at random (MAR) assumption. They then use a re-weighting scheme that follows similar selection function assumptions to those made in this work. In the context of ORB, the possibility to impute not at random by sampling from a different distribution that directly models the selection process could be considered. Additionally, in the PB context, \citet{MI_PB_Carpenter2011} had to impute the study sizes,  which are known in the ORB setting proposed here. Hence, this could be a compelling avenue for future research. In this sense, MI would benefit from modelling outcomes that are correlated to borrow strength in case of missing outcomes \citep{MI_PB_Carpenter2011}, as done in previous work on ORB \citep{Bay, Kirkham2012}. This could be of particular interest in cases such as our motivating example from \citet{Copas2019, topiramate}, where numerous outcomes are considered in the meta-analysis. The challenge in this approach, as previously noted, lies in the estimation of the correlation coefficients \citep{mythesis, Bay, Kirkham2012}.

Our focus was on ORB-adjustment of beneficial outcomes, but the methodology proposed in this work can be easily extended to harmful outcomes by adjusting the selection functions for a different missing data mechanism accordingly. This could mean changing the assumed selection mechanism for unreported outcomes to, for example, assuming that a positive value of the treatment effect for a harmful outcome, or a significant one, results in a lower probability of reporting \citep{Copas2019, Copas2014, ORBIT_paper}. Future implementations of this ORB-adjustment framework could hence investigate which missing data assumptions are reasonable to make for harmful outcomes and, e.g., conduct a simulation study similar to the one done here for various, flexible, selection functions.

For future research on ORB, we encourage the refinement and further exploration of simulation studies and strongly recommend using a pre-defined protocol for transparency and reproducibility. The simulation study conducted in this work utilized a limited range of data-generating mechanism (DGM) parameters and ORB-adjustment selection functions. Future research could involve extensive sensitivity analyses and varying sample sizes to enhance the robustness of the findings, as well as comparisons with new potential approaches, such as the MI ones above-described.

Overall, this study highlights the significant impact of ORB on treatment effect estimation, as well as heterogeneity variance, and demonstrates the efficacy of a flexible ORB-adjustment framework based on selection models. This framework allows the inclusion of contributions from unreported study outcomes and the specification of the desired assumed missing data mechanism via the selection function. The methodology shows promise in mitigating ORB across various settings, with potential for further refinement and broader application.

\section{Code Availability}

The $\texttt{R}$ code for the simulation study presented in this research can be accessed in the \href{https://github.com/agaiasaracini/ORBproject}{ORBproject GitHub repository}. This repository includes all the scripts used in the simulation, featuring the key function $\texttt{reORBadj}$, which implements the ORB-adjustment following the selection model framework described in this study. This function is thus applicable to any meta-analysis dataset with unreported study outcomes that might indicate ORB. By making the simulation study code accessible in the \href{https://github.com/agaiasaracini/ORBproject}{ORBproject GitHub repository} and providing a pre-defined simulation study protocol in the \href{https://osf.io/ancdu/}{OSF project repository}, this research underscores the importance of reproducibility and transparency in scientific investigations.

\bibliographystyle{mywiley} % Choose natbib-compatible bibliography style
\bibliography{biblio} % Replace 'biblio' with your actual BibTeX file name

\newpage

%%%%%%%%%% Bias %%%%%%%%%%%%%%%%%%%%%%%%%%%%%%%%%%%%%%%%%%%%%%%%%%%%%%%%%%%%%%%%

%%%%%%%%%%% MSE %%%%%%%%%%%%%%%%%%%%%%%%%%%%%%%%%%%%%%%%%%%%%%%%%%%%%%%%%%%%%%%%

%%%%%%%%% Coverage %%%%%%%%%%%%%%%%%%%%%%%%%%%%%%%%%%%%%%%%%%%%%%%%%%%%%%%%%%%%%

%%%%%%%%% Empirical SE %%%%%%%%%%%%%%%%%%%%%%%%%%%%%%%%%%%%%%%%%%%%%%%%%%%%%%%%%

%%%%%%%%%%%% Bias tau squared %%%%%%%%%%%%%%%%%%%%%%%%%%%%%%%%%%%%%%%%%%%%%%%%%%

%\bibliographystyle{mywiley} % Choose natbib-compatible bibliography style
%\bibliography{biblio} % Replace 'biblio' with your actual BibTeX file name

\end{document}